\documentclass[12pt]{article}

 \newread\testifexists
 \def\GetIfExists #1 {\immediate\openin\testifexists=#1
     \ifeof\testifexists\immediate\closein\testifexists\else
     \immediate\closein\testifexists\input #1\fi}

 \usepackage{gthstyle}\usepackage{amsfonts}
 \usepackage{amssymb}
 \usepackage{graphicx} \usepackage{epstopdf}
 \immediate\openin\testifexists=e
     \ifeof\testifexists\immediate\closein\testifexists\else
     \immediate\closein\testifexists\input e\fipsf

 \def\Bbb#1{\setbox0=\hbox{$\tt #1$}  \copy0\kern-\wd0\kern .1em\copy0}

 \def\bbf#1{\setbox0=\hbox{$#1$} \kern-.025em\copy0\kern-\wd0
         \kern.05em\copy0\kern-\wd0 \kern-.025em\raise.0433em\box0}

 \immediate\openin\testifexists=a
     \ifeof\testifexists\immediate\closein\testifexists\else
     \immediate\closein\testifexists\input a\fimssym.def  

                     \newcommand{\fn}{\footnote}
              \newcommand{\nm}{\nonumber}
 \newcommand{\be}{\begin{eqnarray}}             \newcommand{\ee}{\end{eqnarray}}
 \newcommand{\bi}[1]{\begin{itemize}\item[#1]}         \newcommand{\itm}[1]{\item[#1]}  \newcommand{\ei}{\end{itemize}}
 \newcommand{\eqn}[1]{(\ref{#1})}

 \newcommand{\crlb}[1]{\label{#1}\\[2pt]}
 \newcommand{\eela}[1]{\quad\hbox{\scriptsize{#1}}\label{#1}\end{eqnarray}}
 \newcommand{\eelb}[1]{\label{#1}\end{eqnarray}}
 
 \newcommand{\newsecb}[2]{\section{#1}\label{#2}\setcounter{equation}{0}}

 \newcommand{\nolabels} {\def\eel{\eelb} \def\crl{\crlb} \def\newsecl{\newsecb}\def\bibiteml{\bibitem}\def\citel{\cite}}

\newcommand\publishversion{\nolabels\setlength{\textheight}{9in}\setlength{\oddsidemargin}{0in}
    \setlength{\textwidth}{6.3in}\setlength{\topmargin}{-0.1in}}

 \def\a{\alpha}      \def\b{\beta}         
 \def\d{\delta}         
 \def\k{\kappa}      \def\l{\lambda}      
 \def\f{\phi}                
 \def\j{\psi}                 \def\s{\sigma}

 \def\pa{\partial} \def\ra{\rightarrow} 
  
 \def\dd{{\rm d}}  \def\bra{\langle}   \def\ket{\rangle}

 \def\qu{\ {\buildrel {\displaystyle ?} \over =}\ }  \def\qqu{\ {\buildrel {\displaystyle ??} \over =}\ }
 
 \def\iss{\ =\ }

 \def\fract#1#2{{\textstyle{#1\over#2}}}
 \def\ffract#1#2{\raise .2 em\hbox{$\scriptstyle#1$}\kern-.3em/
                 \kern-.2em\lower .15 em \hbox{$\scriptstyle#2$}}
 
 \def\half{\fract12} \def\quart{\fract14} 
 
 \def\part#1#2{{\partial#1\over\partial#2}}
 
\def\sgn{\hbox{sgn}}
\def\gl{\,\raisebox{.4em}{$>$}\hspace{-1.1em}\raisebox{-.12em}{$<$}\,}

\publishversion
\begin{document} \begin{titlepage}

\title{\normalsize \hfill ITP-UU-12/14  \\ \hfill SPIN-12/12
\vskip 20mm \Large\bf Relating the quantum mechanics of discrete systems to standard canonical quantum mechanics}

\author{Gerard 't~Hooft}
\date{\normalsize Institute for Theoretical Physics \\
Utrecht University \\ and
\medskip \\ Spinoza Institute \\ Postbox 80.195 \\ 3508 TD Utrecht, the Netherlands \smallskip \\
e-mail: \tt g.thooft@uu.nl \\ internet: \tt
http://www.phys.uu.nl/\~{}thooft/}

\maketitle

\begin{quotation} \noindent {\large\bf Abstract } \medskip \\
Discrete quantum mechanics is here defined to be a quantum theory of wave functions defined on integers \(P_i,\ Q_i\), while canonical quantum mechanics is assumed to be based on wave functions on the real numbers, \({\Bbb R}^n\). We study reversible mappings from the position operators \(q_i\) and their quantum canonical operators \(p_i\) of a canonical theory, onto the discrete, commuting  operators \(P_i\) and \(Q_i\). In this paper we are particularly interested in harmonic oscillators. In the discrete system, these turn into deterministic models, which is our motivation for this study.  We regard the procedure worked out here as a ``canonical formalism" for discrete dynamics, and as a stepping stone to handling discrete deterministic systems in a quantum formalism.
\end{quotation}

\vfill \flushleft{April 19, 2012}
\end{titlepage}
\eject

\def\ol{\overline}  			\def\E{\epsilon}		
\def\qp{\mathrm{qp}}  \def\PQ{\mathrm{PQ}}  \def\edge{\mathrm{edge}}
\def\ds{\displaystyle}	\def\low#1{{\raisebox{-3pt}{\scriptsize{$#1$}}}}
\def\o{\_\!\_\!\_\!}   \def\M{\Box\!}   


\newsecl{Introduction}{intro}
In modern science, real numbers play such a fundamental role that it is difficult to imagine a world without real numbers. Nevertheless, one may suspect that real numbers are nothing but a human invention. By chance, humanity discovered over 2000 years ago that our world can be understood very accurately if we phraze its laws and its symmetries by manipulating real numbers, not only using addition and multiplication, but also subtraction and division, and later of course also the extremely rich mathematical machinery beyond that, manipulations that do not work so well for integers alone, or even more limited quantities such as Boolean variables. 

Now imagine that, in contrast to these appearances,  the real world, at its most fundamental level, were not based on real numbers at all. We here consider systems where only the integers describe what happens at a deeper level. Can one understand \emph{why} our world \emph{appears} to be based on real numbers?

The point we wish to make, and investigate, is that everything we customarily do with real numbers, can be done with integers also. A mapping exists that turns a set of two, mutually commuting,  integer  value  operators, \(P\) and \(Q\), into one real number  valued  operator. We can manipulate this real number anyway we like and subsequently map the result back onto the integers. However,  the  techniques  to  do  this  require the  methods  of  quantum mechanics. Let us first give the general picture.

Regardless what kind of theory we have, as long as its basic dymanical variables are defined to be sets of integers forming a space \({\Bbb Z}^{2n}\), we can always assign an element of a basis of Hilbert space to each point in  \({\Bbb Z}^{2n}\). Having done that, the next thing we can do is transform to a different basis. For every single set of integers \(Q\in\Bbb Z\), we can now consider the basis generated by the functions \(\j_\eta(Q)=\bra Q|\eta\ket=e^{-2\pi i\eta Q}\), normalized by the orthogonality rule \( \bra\eta_1\,|\,\eta_2\ket=\d(\eta_1-\eta_2)\), if both \(\eta_1\) and \(\eta_2\) are chosen to lie within the unit interval \((-\half,\,\half\,]\). This is how any quantum theory on the set of integers can be mapped onto a quantum theory on the set of numbers \(\eta\) on the interval \((-\half,\,\half\,]\), or more precisely, on the unit circle. The mapping is unitary, so we can also have the inverse mapping. The hamiltonian on one side of the mapping is mapped onto a hamiltonian on the other side. 

Subsequently, one might consider deterministic theories. Since these would also either dictate how integers evolve, or how numbers evolve on a circle, these deterministic theories form a subclass of all quantum models on both sides of the mapping, but a deterministic model on one side in general will be mapped onto a pure quantum system on the other. Thus, our work may give further room for speculations about determinism in quantum mechanics, yet in this paper we are mainly interested in the mapping itself.\fn{A separate paper on deterministic versions of a quantum field theory is planned.}

How does the infinite line of real numbers emerge? Simply, we regard a real number as a composite  of its integral part and a fractional part. The integral part, of course, is an integer, and the fractional part lives on the interval  \((-\half,\,\half\,]\). This is the basis of our idea that wave functions defined on the real line can be transformed to functions on two integers, \(P\) and \(Q\), or two circles, \(\eta_1\) and \(\eta_2\). If we know how we wish to transform these to a real coordinate \(q\), we can calculate how its associated momentum \(p\) transforms, since \(p=-i\pa/\pa q\). All we will do now is work out the details, which, at first sight, appear to be straightforward. 

Intuitively, the physical situation may seem to be clear. If we put\fn{For the \(2\pi\), see below under ``notation".} \([q,p]=i/2\pi\),  we have the ``uncertainty relation" \(\d q\cdot\d p\approx 1\); we could take \(\d q\approx 1\) and \(\d p\approx 1\), so that the integers \(Q\) and \(P\) could serve to enumerate a basis. If this is done with some care, the \(P\) and the \(Q\) can be made to commute.

It so happened, however, that there are quite a bit of subtleties and pit falls, forcing us to go slow. To avoid making one of the many possible mistakes, we do our calculations slowly and carefully, which also exposes some beautiful underlying mathematics. 

We refer to this as a \emph{canonical formalism} for discrete dynamics, complementary to the much more familiar canonical formalism of either classical mechanics or quantum mechanics.

\subsection{Notation}\label{notation.sub}

While doing these calculations, the continuous need for factors \(2\pi\) in our numerical expressions and normalization coefficients became irritating. When coordinates and momenta are treated symmetrically, factors \(\sqrt{2\pi}\) appear. A different, but somewhat unusual normalization can simplify things considerably. Instead of writing complex exponentials as \(e^{2\pi i a}\) we will write
	\be e^{2\pi i a}\equiv\E^{i a}\ ;\qquad \E=e^{2\pi}\approx\hbox{ 535{\small .49}}\,\cdots\ ;\qquad \E^{iZ}=1\quad \hbox{if}\quad Z\in\Bbb Z\ . \eel{Exp}
Units of mass, time and energy will be normalized in such a way that\fn{This means that, rather than \(\hbar\),  it is Planck's original constant, \(h\), that is normalized to one.}
	\be 2\pi[x,\,p]=i\ ,\qquad \bra x|\E^{ipa}|p\ket=\bra x+a\,|\,p\ket \ . \eel{commutator}
Throughout the paper, we will use 
	\be\hbox{capital Latin letters,}\qquad&  N,\ P,\ Q,\ X,\ \cdots, &\hbox{ to indicate integers,}\crl{integers}
		\hbox{lower case Latin letters,}\quad&  p,\ q,\ x,\ \cdots, &\hbox{ to indicate  real numbers,}\qquad\crl{real}
		\hbox{and lower case Greek letters,}&  \a,\ \eta,\ \xi,\ \l,\ \cdots, &\hbox{ for fractional numbers, } \eel{Greek} 
the latter being usually confined to the interval \((-\half,\ \half\,] \),  but in expressions that are strictly periodic (with period 1\,), we may for simplicity replace this by the interval \([\,0,\,1)\).

States indicated as \(|\,P\ket\) and \(|\,Q\ket\) form a denumerable basis in Hilbert space. We interpret them as the momenta conjugated to the ``position" operators operators  \(|\,\eta_P\ket,\ |\,\eta_Q\ket\) 
on the periodic unit interval\fn{Due to the unusual normalization of Eqs.~\eqn{Exp}, no factors containing \(2\pi\) are needed here.}:
	\be\bra\eta_Q|\,Q\ket=\E^{iQ\,\eta_Q}\ ,&& \bra\eta_P|\,P\ket=\E^{iP\,\eta_P}\ ,  \crl{orthonormala}
		\bra Q_1|\,Q_2\ket=\d_{Q_1Q_2}\ ,&& \bra\eta^1_Q|\,\eta^2_Q\ket=\d(\eta^1_Q-\eta^2_Q)\ , \crl{orthonormalb}
		\hbox{and on the real line:}\quad \bra x|p\ket=\E^{ipx}\ ,&&\!\!\bra x_1|x_2\ket=\d(x_1-x_2)\ ,\quad\bra p_1|p_2\ket=\d(p_1-p_2)\ .\nm\ee
Furthermore, the operator \(\eta_Q\) obeys: \(\E^{i\eta_QN}|Q\ket=	|Q+N\ket\).\quad On the unit interval \((-\half,\,\half)\), we can Fourier expand
	\be \eta =\sum_{N}\E^{iN\eta} \int_{-\half}^{\half}\eta_1\dd\eta_1\, \E^{-iN\eta_1}\iss\sum_{N\ne0} {i(-1)^N\over 2\pi N} \,\E^{iN\eta}\ ,\hbox{ so that the \(\ \eta_Q\ \) operator obeys} \quad \nm\ee
	\be \eta_Q = \sum_{N\ne 0}{i\over 2\pi N}(-1)^N\E^{iN\eta_Q}\ ,\qquad \bra Q_1|\eta_Q|Q_2\ket={i\over 2\pi}(\d_{Q_1Q_2}-1){(-1)^{Q_2-Q_1}\over Q_2-Q_1}\ . \eel {etaQelements}
The commutator between the operators \(\eta_Q\) and \(Q\) is therefore not quite what one might have expected:
	\be\bra Q_1|[\eta_Q,\,Q]|Q_2\ket =(Q_2-Q_1)\bra Q_1|\eta_Q|Q_2\ket={i\over 2\pi}\left(\d_{Q_1Q_2}-  (-1)^{Q_2-Q_1}\right)\ .\eel{etaQcomm}
The extra term can be written as \(-\fract i{2\pi}|\j_1\ket\bra \j_1|\) where the state \(|\j_1\ket\), defined by \(\bra Q|\j_1\ket=(-1)^Q\), sits at the edge of the \(\eta_Q\) interval. It arises because the \(\eta_Q\) operator is \emph{not} itself periodic, unless it is forced to jump back by one unit across the point \(\eta=\pm\half\), and this leads to a Dirac delta function in its derivative. We will refer to this state as an \emph{edge state}. It will cause us quite some trouble, but, as we shall see, the edge states can be tamed.
	
\newsecl{A first attempt}{attempt.sub}
Let us start doing what we intended to do from the very beginning: identify a position operator \(q\) as
		\be q\qu Q+\eta_P\ ;\qquad |q\ket\equiv|Q,\,\eta_P\ket\ .\eel{qoperator}
This is totally legal, in principle; since \(P\) and \(Q\) are different integers, we work in the product Hilbert space, so that \(Q\) and \(\eta_P\) are commuting operators. All states \(|q\ket\) are represented exactly once, so that this identification is invertible. The question mark here indicates that we will replace this expression by a better choice later.
	
Computation of the associated \(|p\ket\) states is straightforward:
	\be|p\ket=\int_{-\infty}^{\infty}\dd q|q\ket\bra q|p\ket=\int_{-\infty}^{\infty}\dd q\,\E^{iqp}|q\ket\qu\sum_Q\int_{-\half}^{\half}\dd\eta_P\,\E^{ip(Q+\eta_P)}|Q,\eta_P\ket\ . \eel{pstate}
Let us write \(p=K+\k\), where, in line with the notation that we will always use, \(K\) is an integer and \(|\k|\le\half\).
	Now we have the general integral expression
	 \be \int_{-\half}^{\half}\dd \eta\, \E^{i( N+\k)\eta}={2\sin\pi\k\over 2\pi}\,{ (-1)^N\over N+\k}\ ,\qquad N\hbox{ integer,}\quad\k\hbox{ fractional,} \eel{int1}
(which,  In the limit \(\k\ra0\),  turns into \(\d_N\)), and from this we find
	\be \bra Q_1,P_1|K+\k\ket\ \qu\ {\sin\pi\k\over \pi}\,{(-1)^{K-P_1}\over K-P_1+\k}\,\E^{i\k Q_1}\ , && \crl{momstate}
\hbox{ while from \eqn{qoperator}, we have: }\qquad\qquad\bra Q_1,P_1|\!\:Q+\eta_P\ket\ \qu\ \d_{QQ_1}\E^{-i\eta_P\,P_1}\ .&&\qquad \eel{posstate}
For future use, we also need the corresponding expressions in \(\eta_Q,\eta_P\) space. One easily derives
\be\bra\eta_Q^1,\eta_P^1|K+\k\ket&\qu&\d(\eta_Q^1+\k)\E^{i\eta_P^1(K+\k)}\ .		\crl{etaspacemom}
	\bra\eta_Q^1,\eta_P^1|Q+\xi\ket&\qu&\d(\eta_P^1-\xi)\E^{i\eta_Q^1Q}\ .		\eel{etaspacepos}

A drawback may seem to be that our treatment of the operators \(p\) and \(q\) is asymmetric, but this is easy to remedy; we can replace Eqs.~\eqn{etaspacemom} and \eqn{etaspacepos} by
	\be\bra\eta_Q^1,\eta_P^1|K+\k\ket&\qqu&\d(\eta_Q^1+\k)\E^{i\eta_P^1(K+\s\k)}\ ,		\crl{etasymmom}
	\bra\eta_Q^1,\eta_P^1|Q+\xi\ket&\qqu&\d(\eta_P^1-\xi)\E^{i\eta_Q^1(Q+\ol\s\xi)}\ ,		\eel{etasympos}
and verify that the first parts of Eq.~\eqn{pstate} still hold if the parameters \(\s\) and \(\ol\s\) obey
	\be\s+\ol\s=1\ .\eel{sigmasum}
We obtain
	\be \bra Q_1,P_1|K+\k\ket&\qqu&{2\sin(\pi\s\k)\over 2\pi}\,{(-1)^{K-P_1}\over K-P_1+\s\k}\,\E^{i\k Q_1}\ , \crl{sigmamom}
             \bra Q_1,P_1|Q+\xi\ket&\qqu&{2\sin(\pi\ol\s\xi)\over 2\pi}\,{(-1)^{Q-Q_1}\over Q-Q_1+\ol\s\xi}\,\E^{-i\xi P_1}\ , 
             		\eel{sigmapos}
and symmetry is obtained when \(\s=\ol\s=\half\).

One might have hoped for a formalism where   \(K\) coincides with \(P_1\) and \(Q\) with \(Q_1\), but we see that this cannot be made generally true. The kernels, which vanish no faster than \(1/|K-P_1|\) or \(1/|Q-Q_1|\), give problems when we wish to construct the operators \(p\) and \(q\) themselves: non-canonical terms appear in their commutator. Again, all this is due to edge states. One way to avoid disaster is by only using soft operators such as 
\(\E^{i\a p}\) and \(\E^{i\b q}\), to be constructed by using
	\be \bra Q_1,P_1|\E^{i\a p}| Q_2,P_2\ket\equiv\sum_K\int_{-\half}^{\half}\dd\k\bra Q_1,P_1|K+\k\ket\E^{i\a(K+\k)}\bra K+\k|Q_2,P_2\ket\ , \eel{expmomop} and so on, since these sums and integrals converge well. Indeed, one then obtains the desired commutation rule
		\be \E^{i\a p}\,\E^{i\b x}=\E^{i\a\b}\,\E^{i\b x}\,\E^{i\a p}\ . \ee
This however does not allow us to compute hamiltonians and other interesting operators directly in \((P,\,Q)\) space, which is why a superior procedure was searched for. It turns out that all of the above definitions for the \(p\) and \(q\) states can be replaced by better expressions, which is why we adorned them with question marks.

\newsecl{(Almost) removing the edge states}{almost}
If the \(p\) and \(q\) states of the previous section are used to compute the commutator \([q,p]\), one finds that, at the edges of the intervals where \(\eta_Q\) and \(\eta_P\) are defined, the edge states give non-canonical contributions. Although there is nothing wrong with defining our mappings this way, it turns out that the contributions of the edge states are cumbersome. It is better to reduce their effects to a minimum.To achieve this, we have to recover full periodicity in \(\eta_Q\) and \(\eta_P\).  For technical reasons, we rename these as \(\eta_1\) and \(\eta_2\).

Consider the torus defined by the intervals \(-\half<\eta_1\le \half\) and \(-\half<\eta_2\le\half\). In order to avoid all contributions from edge states, we need to define the operators \(q\) and \(p\) unambiguously on this torus, with full periodicity. From the expressions  \eqn{etasymmom}--\eqn{sigmapos}, we are led to try\fn{Later we will observe that no generality is lost if we return to the case \(\s=1,\ \ol\s=0\), but keeping the two parameters \(\s,\,\ol\s\) makes it easier to observe the \( p\leftrightarrow q \) symmetry}:
	\be q=-\fract i{2\pi}\pa_1+\s\eta_2\ ,\quad p=-\fract i{2\pi}\pa_2-\ol\s\eta_1\ ;\qquad \pa_1=\pa/\pa\eta_1\ ,\quad\pa_2=\pa/\pa\eta_2\ , \eel{qpdef}
with \(\s+\ol\s=1\). Indeed, since
	\be [\pa_i,\eta_j]=\d_{ij}\ ,\quad[\eta_1,\,\eta_2]=0\ ,\quad\hbox{and}\quad[\pa_1,\,\pa_2]=0\ , \eel{etacomrules}
we find that Eq.~\eqn{qpdef} ensures that \([q,\,p]=\fract i{2\pi}\). \quad However, if we also desire that \(q=Q+\eta_2\) and \(p=K-\eta_1\), where \(Q\) and \(K\) are always integral then the wave functions \(\j_\qp(\eta_1,\eta_2)\) allowed on the torus must always obey the following continuity properties across the borders:
	\be\j_\qp(+\half,\,\eta_2)=\j_\qp(-\half,\,\eta_2)\E^{i\ol\s\eta_2}\ ;\qquad\j_\qp(\eta_1,\,+\half)=\j_\qp(\eta_1,\,-\half)\E^{-i\s\eta_1}\ . \eel{etaperiods}
The \(q\) and \(p\) eigenstates are then:
	\be\hbox{If }\ q=X+\xi\ : &&\bra\eta_1,\eta_2|q\ket = \d(\eta_2-\xi)\E^{iX\eta_1+i\ol\s\eta_1\eta_2}\ ,\qquad \hbox{and}\crl{qeigen}
	      \hbox{If }\ p=K+\k\ :  &&\bra\eta_1,\eta_2|p\ket= \d(\eta_1+\k)\E^{iK\eta_2-i\s\eta_1\eta_2}\ . \eel{peigen}

Now, observe something that is of crucial importance: the only way to introduce on the torus the operators \(Q=-i\pa_1\) and \(P=-i\pa_2\), where both \(Q\) and \(P\) are restricted to be integers, is to have \emph{regular} periodicity. We need a Hilbert space of wave functions \(\j_\PQ(\eta_1,\eta_2)\) that obey:
	\be\j_\PQ(+\half,\,\eta_2)=\j_\PQ(-\half,\,\eta_2)\ ;\qquad\j_\PQ(\eta_1,\,+\half)=\j_\PQ(\eta_1,\,-\half)\ . \eel{etaregperiods}
To relate the functions \(\j_\qp\) to \(\j_\PQ\), we can either introduce a factor \(\E^{i\ol\s\eta_1\eta_2}\), so that the first part of Eq.~\eqn{etaperiods} agrees with the first part of Eq.~\eqn{etaregperiods}, or a factor \(\E^{-i\s\eta_1\eta_2}\) to make the seconds parts agree, but to make them all agree is harder. The interpolating function on the torus must be unimodular if we want the transformation to be unitary, while its phase must make a full rotation over \(2\pi\) while circulating around the edge if the unit square --- in short: \emph{the interpolating factor must contain one unit of phase flux}. This gives the interpolating function a singularity, which is easiest to situate on the corner: \((\eta_1=\pm\half,\,\eta_2=\pm\half)\). The required properties of this phase function \(\E^{i\f(\eta_1,\eta_2)}\) are illustrated in Fig.~\ref{philines.fig}.

\begin{figure}[h] \setcounter{figure}{0}
\includegraphics[width=90 mm]{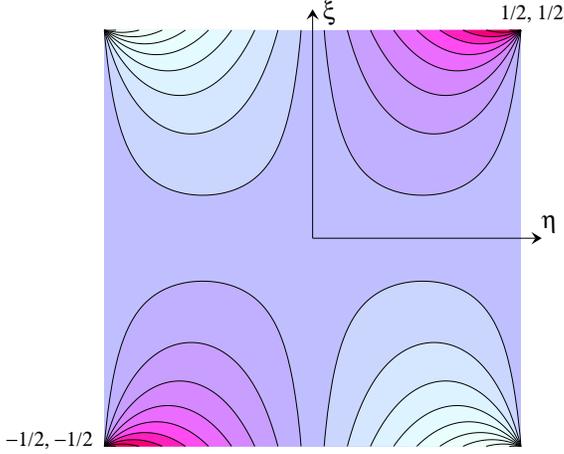}
 \begin{center}   \caption{An artist's impression of the contour lines of the function \(\f(\eta,\xi)\), defined in Eq.~\eqn{phidef}. Its values run from \(-\quart\) (top left) 
 to \(+\quart\) (top right), corresponding to the angles \(-90^\circ\) to \(90^\circ\).}\label{philines.fig}
\end{center}
\end{figure}

With such an interpolating phase function, we can map the real line on the double set of  integers \((Q,\,P)\). The fact that there is a singularity means that there is one edge state left:
	\be \bra\eta_1,\eta_2|\j_\edge\ket=\d(\eta_1-\half)\d(\eta_2-\half)\ ,\qquad   \bra Q,\,P\,|\j_\edge\ket=(-1)^{P-Q}\ . \eel{singleedge}
This has to be contrasted with the situation we had before: there was an edge state \(\bra Q\,|\j_1\ket=(-1)^Q\) for all values of \(P\) and an edge state \(\bra P\,|\j_2\ket=(-1)^P\) for all values of \(Q\). We now have just a single edge state left. This improves convergence sufficiently to allow us to move on and study the \(q\) and \(p\) states in the \((Q,P)\) Hilbert space without further trouble. The single edge state cannot be avoided since it is associated with a conserved flux on the torus. 

We do have to construct explicit expressions for the interpolating function. The desired function is constructed in two steps. First, find a smooth complex function \(f(\eta_1,\eta_2)\) obeying both boundary condtions \eqn{etaperiods}. A suitable choice could be:
	\be f(\eta_1,\eta_2)\qu\E^{-i\ol\s\eta_1\eta_2}\cos\pi\eta_2+\E^{i\s\eta_1\eta_2}\cos\pi\eta_1 \ , \eel{interpoltrial}
and we can confirm here that this function can serve our purpose quite well. It can be used to connect functions with the matching conditions \eqn{etaperiods} with functions obeying \eqn{etaregperiods}. However, it is not the optimal choice, since the border crossings are not completely analytic. We found a better solution in terms of the elliptic theta functions\cite{GR}. We choose:
	\be f(\eta_1,\eta_2)=\E^{-i\overline\s\eta_1\eta_2}\sum_{N=-\infty}^\infty\E^{-\half N^2+N(i\eta_1+\eta_2)}\ .\eel {interpol}
Within the intervals \(|\eta_1|\le \half\) and \(|\eta_2|\le\half\), the functions \eqn{interpoltrial} and \eqn{interpol} are so similar that their effects are very nearly the same, but they do not coincide.

 	The interpolating function is then
	\be U(\eta_1,\eta_2)={f(\eta_1,\eta_2)\over |f(\eta_1,\eta_2)|}\ ;\qquad
	\j_\PQ(\eta_1,\eta_2)=U(\eta_1,\eta_2)\, \j_\qp(\eta_1,\eta_2)\ . 
	\eel{Uinterpol}
Note that, if \(\eta_1=\half\)+a and \(\eta_2=\half\)+b, with \(a\) and \(b\) infinitesimal, then \(\E^{i\ol\s\eta_1\eta_2}/\E^{-i\s\eta_1\eta_2}=i\), while \(\cos\pi\eta_1=-\pi a\) and \(\cos\pi\eta_2=-\pi b\), so the function \(f\) in Eq.~\eqn{interpoltrial} makes a full phase rotation if we follow a small curve around the point \((\half,\half)\). The function in Eq.~\eqn{interpol} does the same (See Appendix \ref{prop}). Both functions have a zero in the corners \((\pm\half,\,\pm\half)\).

The operators \((\eta_Q,\eta_P)\) are now defined to take the same values as \((\eta_1,\eta_2)\), but we define their joint fundamental eigen states to be rotated by a  
factor \(U(\eta_1,\eta_2)\). Thus, we rewrite \eqn{Uinterpol} as follows:
	\be|\eta_Q,\eta_P\ket\equiv U^*(\eta_1,\eta_2)\,|\eta_1,\eta_2\ket\ . \eel{rotate}
Thus, Eqs.~\eqn{qeigen} and \eqn{peigen} for the \(p\) and \(q\) eigen states are now written as
	\be\bra\eta_Q,\eta_P|q\ket &=& U (\eta_Q,\eta_P)\,\d(\eta_P-\xi)\E^{iX\eta_Q+i\ol\s\eta_Q\eta_P}\ ,\quad q=X+\xi\ ;\qquad \hbox{and}\crl{qeta}
		\bra\eta_Q,\eta_P|p\ket &=& U (\eta_Q,\eta_P)\,\d(\eta_Q+\k)\E^{iK\eta_P-i\s\eta_Q\eta_P}\ ,\quad p=K+\k\ . \eel{peta}
Note that \(\ol\s\) (\(\s\)) dependence in Eqs.~\eqn{qeta}, \eqn{peta}, \eqn{interpoltrial} and \eqn{interpol} cancels out. We had kept \(\ol\s\) and \(\s\) just to demonstrate the logical coherence of these equations. Clearly, the \(p\leftrightarrow q\) symmetry is fully maintained in our present procedure.	

Writing\fn{We apologise for the use of brackets ( ) that can mean two different things; most often they are just meant to group terms together when multiplied, 
but in expressions such as  \(\f(\eta,\xi)\) they indicate that \(\f\) is a function of \(\eta\) and \(\xi\). The comma should make this unambiguous.}:
 	\be \sum_N\E^{-\half N^2+N(i\eta+\xi)}=r(\eta, \xi)\,\E^{i\f(\eta,\xi)}\ , \eel{phidef}
with \(r\) and \(\f\) real, we need the phase function \(\f(\eta,\xi)\). Its most important properties are that it is differentiable, and, when using \eqn{interpol}, we also have 
	\be&\f(\eta,\xi+1)=\f(\eta,\xi)+\eta  ;\qquad\f(\eta+1,\xi)=\f(\eta,\xi)\ ;& \crl{pseudoperiods}
	&\f(\eta,\xi) = -\f(-\eta,\xi) =  -\f(\eta,-\xi)\ . &\eel{phiproperties}
One can also prove the important relation:
	\be\f(\eta,\xi)+\f(\xi,\eta)=\xi\eta\ , \eel{phisymsum}
and in Eq.~\eqn{rotate} we use:
	\be U(\eta_1,\eta_2)=\E^{i\ol\s\eta_1\eta_2-i\f(\eta_1,\eta_2)}=\E^{-i\s\eta_1\eta_2+i\f(\eta_2,\eta_1)}\ . \eel{Uphi}

Eqs.~\eqn{qeta} and \eqn{peta} have now become
	\be\bra\eta_Q,\eta_P|q\ket=\E^{iX\eta_Q+i\f(\eta_Q,\xi)}\,\d(\eta_P-\xi)\ ;\quad\bra\eta_Q,\eta_P|p\ket=\E^{iK\eta_P+i\f(\eta_P,\k)}
		\,\d(\eta_Q+\k)\ . \eel{qpphi}
Since, in these equations, the states \(|q\ket\) and \(|p\ket\) are now fully periodic in \(\eta_P\) and \(\eta_Q\), it finally became legitimate to expand in the basis 
\(|Q,P\ket=\E^{iP\eta_P+iQ\eta_Q}|\eta_Q,\eta_P\ket\), to obtain
	\be\bra Q,P|q\ket=\E^{-iP\xi}\int_{-\half}^\half\dd\eta\,\E^{i\f(\eta,\xi)+i(X-Q)\eta}\ ;\quad\bra Q,P|p\ket=\E^{iQ\k}\int_{-\half}^\half\dd\eta\,\E^{i\f(\eta,\k)+i(K-P)\eta}\ . \eel{pqphi}
 This expression for \(\bra 0,0|q\ket\) can be inverted to give
	\be \E^{i\f(\eta,\xi)}=\sum_X\bra 0,0|X+\xi\ket\,\E^{-iX\eta}\ . \eel{phiq}

\begin{figure}[h] 
 \includegraphics[width=145 mm]{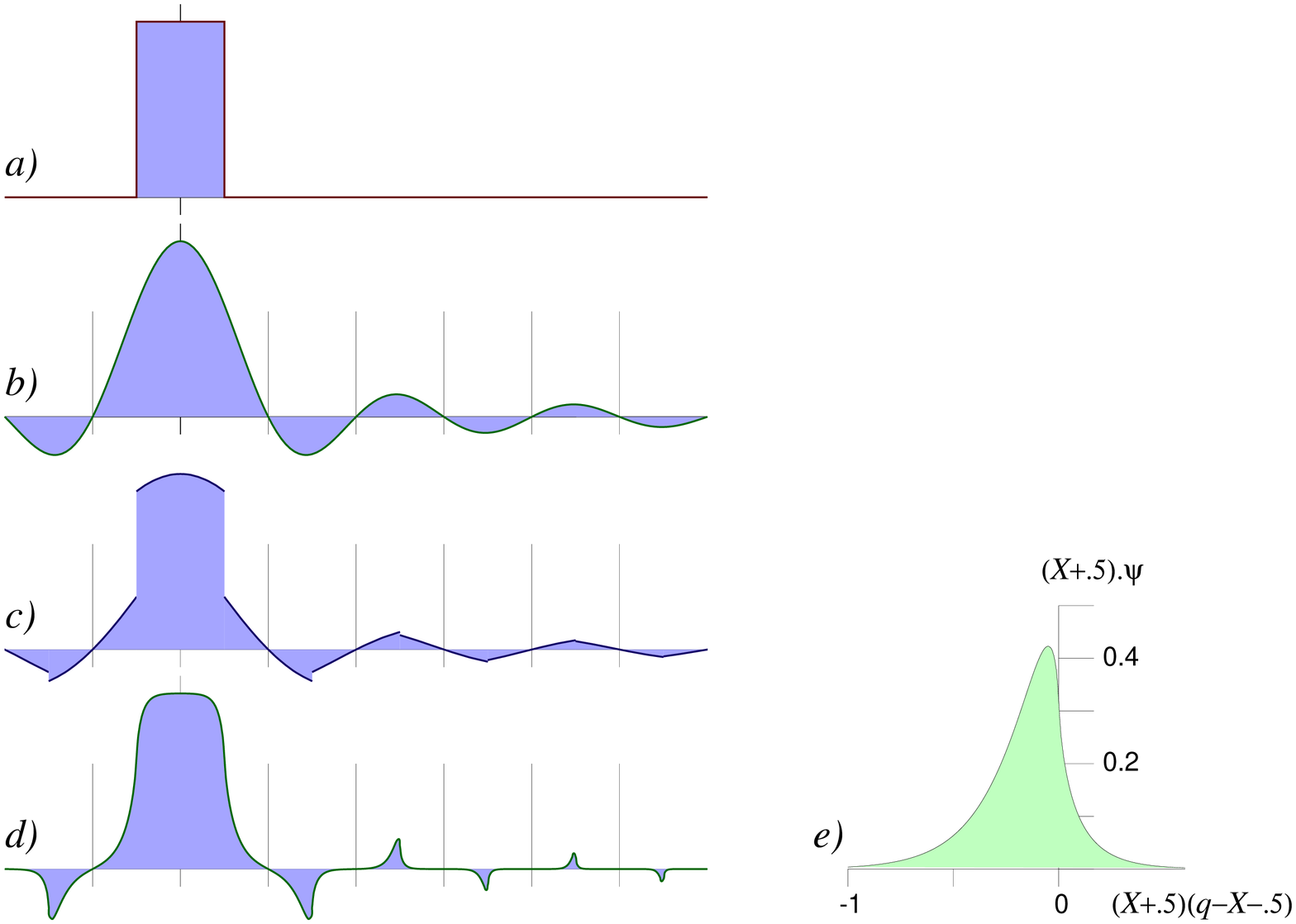}
  \caption{\(|Q,P\ket\) states in various cases. $a)$ Our starting point, he state \(|0,0\ket\) in \(q\) space, Eq.~\eqn{posstate} . $b)$ Its Fourier transform, in \(p\) space, Eq.~\eqn{momstate}, converges only slowly for large \(p\). $c)$ The \(|0,0\ket\) state according to Eq.~\eqn{sigmapos}. It is equal to its own Fourier transform (its form in \(p\) space). $d)$ The \(|0,0\ket\) state according to Eq.~\eqn{pqphi}. It is also equal to its Fourier transform in \(p\) space, and converges better. The other \(|Q,P\ket\) states are obtained from these by translation and/or multiplication with \(\E^{iP\:\!q}\).\quad $e)$ The small peaks in the curve $d$ quickly reach this asymptotic form, if multiplied both horizontally and vertically with a factor \((X+.5)\). The surface areas of these peaks rapidly approach the values \(\pm 1/(2\pi(X+\half)^2)\).} \label{states.fig}
\end{figure}

To illustrate what was gained by this new mapping between \(|q\ket\) and \(|p\ket\) states on the one hand and the \(|Q,P\ket\) states on the other, the wave functions \(\bra q|0,0\ket\) and \(\bra p|0,0\ket\) in various schemes are illustrated in Fig.~\ref{states.fig}. The Figure suggests that the function \(\bra 0,0|q\ket\), for large values of \(|q|\), tends to
	\be\sum_X{(-1)^X\d(\,|q|\,-X-\half)\over2\pi(X+\half)^2}\ . \eel{asymptdelta}
The matrix elements \(\bra Q,P|q\ket\) and \(\bra Q,P|p\ket\) can all be easily expressed in terms of the one basic function  \(\j(q)=\bra 0,0|q\ket\). Further properties of this function are expanded upon in Appendix~\ref{prop}.

The contour lines of the function \(\f(\eta,\xi)\) are sketched in Fig.~\ref{philines.fig}. Observe the singularity due to the edge state in the corners.
In \((\eta_Q,\eta_P)\) space, the \(q\) and \(p\)-operators now read:
	\be q&=&{-i\over 2\pi}{\pa\over\pa\eta_Q}+\left({\pa\over\pa\eta_Q}\f(\eta_P,\eta_Q)\right)\iss
				{-i\over 2\pi}{\pa\over\pa\eta_Q}+\eta_P-\left({\pa\over\pa\eta_Q}\f(\eta_Q,\eta_P)\right)\ ; \crl{qopetaspace}
		p&=&{-i\over 2\pi}{\pa\over\pa\eta_P}-\left({\pa\over\pa\eta_P}\f(\eta_Q,\eta_P)\right)\iss
				{-i\over 2\pi}{\pa\over\pa\eta_P}-\eta_Q+\left({\pa\over\pa\eta_P}\f(\eta_P,\eta_Q)\right)	\ , \eel{popetaspace}
where we used Eq.~\eqn{phisymsum}.		
Note that, according to Eqs.~\eqn{phiproperties}, these expressions are exactly periodic in both \(\eta_Q\) and \(\eta_P\), with the 
only singularities being in the corners of the \((\eta_Q,\eta_P)\)-quadrant. Apart from possible effects due to this singularity, we have,
everywhere in \((\eta_Q,\eta_P)\) space,
	\be[q,p]=i/2\pi\ . \eel{qpcom}
What was gained by adding the \(\f\) field is, that there are no contributions from \(\d\) functions on the boundaries of our quadrant. Nevetheless, the effects of the one remaining edge state
still show up in the commutator \eqn{qpcom} when expressed in the \(Q,P\) basis. The relevant matrix elements are computed in Appendix~\ref{matrix}.

\newsecl{The harmonic oscillator}{harm}
	We are interested in the fate of the quantum harmonic oscillator in \(p\)-\(q\) space when we map it to the discrete \(P,Q\) variables. 
To find out, we first have to go the \(\eta_Q,\eta_P\) quadrant. In our units, the hamiltonian is
	\be H=\pi(p^2+q^2)=a^\dag\;\! a+\half\ ;\qquad \fract{\pa}{\pa t}\j=-2\pi i H\j\ ,\qquad\j(t)=\E^{-iHt}\j(0)\ . \eel{Schroedinger}
Creation and annihilation operators are
	\be a^\dag=\sqrt\pi(p+ iq)\ ,\quad\ a=\sqrt\pi(p- iq)\ ;\qquad [a,a^\dag]=1\ . \ee
The eigen states \(|\j_n\ket\), \(n=0,1,\cdots\), are given by
	\be a|\j_0\ket=0\ ,\qquad |\j_n\ket=\fract 1{\sqrt{n!}}a^{\dag\,n}|\j_0\ket\ . \eel{harmstates}
In terms of \(\eta_Q\) and \(\eta_P\), for simplicity again written as \(\eta_1,\eta_2\), the annihilation operator is	
	\be a=\sqrt\pi \left(-\fract 1{2\pi}(\pa_1+i\pa_2)-i\eta_2+\fract i{2\pi}(\pa_1+i\pa_2)\f(\eta_1,\eta_2)\right)\ . \eel{annihileta}

Solving the differential equation \(a|\j_0\ket=0\) appears to be easy. Introduce \(z_\pm=\eta_1\pm i\eta_2\), to find
	\be \pa_-(|\j_0\ket\E^{\half\eta_2^2 })=i\pa_-\f(\eta_1,\eta_2)|\j_0\ket\ ;\qquad\bra\eta_1,\eta_2|\j_0\ket=f(z_+)\E^{i\f(\eta_1,\eta_2)-\half\eta_2^2	}\ .\eel{fz}
The function \(f(z_+)=f(\eta_1+i\eta_2)\) is now determined by the boundary conditions \eqn{phiproperties}. The term \(\half\eta_2^2\) plays no role there, because
it matches periodic boundary conditions, but the behavior of \(\f\) is more difficult to accommodate for. Fortunately, of course, we know the solution in \(q\) space:
	\be\bra q|\j_0\ket=2^{1/4}	\E^{-\half q^2}\ , \eel{psinulq}
So, using Eq.~\eqn{qpphi},
	\be\bra\eta_1,\eta_2|\j_0\ket&=&2^{1/4}\sum_{X=-\infty}^\infty\int_{-\half}^\half\dd\xi\d(\eta_2-\xi)\E^{-\half(X+\xi)^2+iX\eta_1+i\f(\eta_1,\xi)}\ = \\
	&=& 2^{1/4}\E^{i\f(\eta_1,\eta_2)-\half\eta_2^2}\sum_X\E^{-\half X^2+iX(\eta_1+i\eta_2)}\ , \eel{groundeta}
which is indeed of the form \eqn{fz}.
		
The expression \eqn{groundeta} could have been arrived at directly by the following chain of arguments: the periodicity requirements \eqn{phiproperties} imply strict periodicity in	
\(\eta_1\); therefore, the wave function can be expanded in waves periodic in \(\eta_1\): \(f(z_+)= \sum_{X\in \Bbb Z}\, a\low{X}(\eta_2)\,\E^{iX\eta_1}\). This must be a function of \(\eta_1+i\eta_2\); therefore,
	\be f(z_+)=\sum_X a\low{X}\,\E^{iX(\eta_1+i\eta_2)}\ . \eel{deriveground}
According to the first equation \eqn{phiproperties}, we must require
	\be f(\eta_1+\half i)&=&f(\eta_1-\half i)\E^{-i\eta_1}\ ;\qquad\hbox{therefore,}\ee
	\be	\sum_X a\low{X}\,\E^{iX\eta_1-\half X}&=&\sum_Xa\low{X}\E^{iX\eta_1+\half X-i\eta_1}\iss \sum_X a\low{X}\E^{i(X-1)\eta_1+\half X} \\
			&=&\sum_Xa\low{X+1}\,\E^{iX\eta_1+\half(X+1)}\ ,\qquad\hbox{or} \\
			a\low{X+1}&=&a\low{X}\,\E^{-X-\half}\ ;\qquad a\low X=C\E^{-\half X^2}\ . \ee

To construct the \(Q,P\) matrix elements of the hamiltonian \eqn{Schroedinger}, it seems to be best to use the wave functions of Fig.~\ref{states.fig}$d$,
	\be\bra q|Q,P\ket=\E^{iPq}\,\bra q-Q|0,0\ket \qquad\hbox{and}\qquad \bra p|Q,P\ket=\E^{-iQp}\,\bra p-P|0,0\ket\ ,\eel{QPqp}
as templates --- note, that the functions \(\bra q|0,0\ket\) and \(\bra p|0,0\ket\) are given by the same mathematical expression.  Then the matrix elements of \(H=\pi(q^2+p^2)\) are found by calculating
	\be\bra Q_1,P_1|q^2|Q_2,P_2\ket&=&\int_{-\infty}^\infty \dd q\,\E^{iq(P_2-P_1)}\bra 0,0|q-Q_1\ket q^2\bra q-Q_2|0,0\ket\ ;\crl{qqelements}
		\bra Q_1,P_1|p^2|Q_2,P_2\ket&=&\int_{-\infty}^\infty \dd p\,\E^{ip(Q_1-Q_2)}\bra 0,0|p-P_1\ket p^2\bra p-P_2|0,0\ket\ , \eel{ppelements}
where the curves of Fig.~\ref{states.fig} must be inserted. Considering the asymptotic form \eqn{asymptdelta}, we notice that both integrals here diverge logarithmically. This requires a logarithmically  infinite subtraction in these matrix elements, proportional to \((-1)^{Q_1-Q_2+P_1-P_2}\), which is exactly the contribution of the one remaining edge state, Eq.~\eqn{singleedge}. It is a harmless additive constant in \(H\).

The hamiltonian \eqn{Schroedinger} describes a harmonic oscillator with period \(T=1\), but in terms of the discrete states \(|P,Q\ket\), the evolution is \emph{deterministic} over multiples of one quarter of this period, as will be demonstrated now. As is well-known for a harmonic oscilator, it sends position states into momentum states and vice versa\fn{In fact, the harmonic oscillator is the ideal instrument to produce \emph{fractional Fourier transforms}\cite{FRFT}, by considering how a wave function transforms at arbitrary, fractional time \(t\).}. If at \(t=0\) we have a state
	\be\bra q|\j_{t=0}\ket=f(q)\ , \eel{initialharmonic}
then at one quarter of a period one finds
	\be\bra p|\j_{t=1/4}\ket=f(-p)\ . \eel{quartharmonic}
and so on. So now, because of the symmetry in the states \(|p\ket\) and \(|q\ket\) in the expressions \eqn{QPqp}, we find that if the initial state is
	\be\bra Q,P|\j_{t=0}\ket=\d_{QA}\,\d_{PB}\ , \eel{initialdiscr}
then after one quarter of a period,
	\be\bra Q,P|\j_{t=1/4}\ket=\d_{QB}\,\d_{P,-A}\ , \eel{quartdiscr}
and so on. This is a kind of determinism that we plan to study further: \(Q\) turns into \(P\) and \(P\) turns into \(-Q\).	What is new in our formalism is that we identified a hamiltonian that does this job in \(Q,P\) space, while it has a natural and non-trivial ground state, Eq.~\eqn{groundeta}.

\newsecl{The edge state}{edge}
The one edge state of our system is  \(\j_\edge\), described by 
	\be \bra\eta_Q,\eta_P|\j_\edge\ket=\d(\eta_Q-\half)\,\d(\eta_P-\half)\quad\hbox{or}\quad\bra Q,P|\j_\edge\ket=(-1)^{P+Q}\ . \eel{edgestate}
Here, we did not attempt to normalize it; the norm squared of this state would be \(L_Q L_P\), if we would have had a box of length \(L_Q\) in the \(Q\) direction and \(L_P\) in the \(P\) direction. It is this state that causes a logarithmic divergence in the hamiltonian of a harmanic oscillator, but more importantly, it generates a non-canonical term in the commutation rule for the real number operators \(q\) and \(p\) that we constructed. In Eq.~\eqn{pqcomm} in Appendix \ref{matrix}, we see that the non-canonical contribution to this commutator is proportional to
	\be|\j_\edge\ket\bra\j_\edge|\ , \eel{edgeproject}

Thus, to achieve a useful mapping with a standard canonical system, somehow, this state must be filtered out. It would be tempting to argue that we should simply replace all states \(|\j\ket\) by
	\be |\j_{\mathrm{phys}}\ket\qu(\,1-\,\fract 1{L_QL_P}\,|\j_\edge\ket\bra\j_\edge|\,)\j\ket\ , \eel{physstate}
but this is not the entire story. When acting with operators such as \(q\) and \(p\) on such a ``physical" state, a non-physical state might emerge. A non-physical state is a state \(|\j\ket\) that obeys
	\be \sum_{P,Q}(-1)^{Q+P}\bra Q,P|\j\ket\equiv a(\j)\ne 0\ . \eel{nonphysstate}
	
There is a much better way to look at the edge state. To see how to remove an unphysical state correctly, we now formulate an important theorem:
\begin{quotation} \noindent Let \(|\j\ket\) be a state with the following properties:
	\bi{1.} It has a \emph{compact support} in \((Q,P)\) space, or equivalently, for only a finite number of values \(Q\) and \(P\), its components \(\bra Q,P|\j\ket\ne 0\), and
	\itm{2.} Its ``edge state coefficient" \(a(\j)\), as defined in Eq.~\eqn{nonphysstate}, is not equal to zero, \ei
	then the operator \(p^2\) acting on this state has a positive, infinitely large expectation value: \(\bra\j|p^2|\j\ket\ra\infty\).
\end{quotation}
To prove this theorem, just consider the matrix elements \eqn{pmatrix}, \eqn{aPmatrix} of \(p\), derived in Appendix~\ref{matrix}, and use them to derive the asymptotic form of the matrix elements 
	\be \bra Q,P|\,p\,|\j\ket=\sum_{Q_2,P_2}\bra Q,P|p|Q_2,P_2\ket\bra Q_2,P_2|\j\ket\ , \eel{ppsi}
for large values of \(Q\) and \(P\). Because all values of \(Q_2\) and \(P_2\) are bounded, we find that only the ``vector potential" operator \(a_P\) contributes for large \(Q\) and \(P\), and these \(Q,P\) matrix elements approach the values 
	\be {\pm iQ\,a(\j)\over P^2+Q^2}\ . \eel{ppsiasympt}
Therefore, the expectation value of \(p^2\) receives the asymptotic contribution
	\be \bra\,p^2\,\ket\ra\sum_{Q,P}{Q^2|a(\j)|^2\over(P^2+Q^2)^2}\ , \eel{divsum}
and the sum of these positive terms diverges logarithmically. \emph{Q.E.D.}
	
Now, we use this result as follows. \emph{If} we decide to map our discrete system on a continuous model where the hamiltonian has the form 
	\be H\approx \fract 1{2m}\,p^2+V(q)\ , \eel{pqham}
where \(V(q)\) is bounded from below, then any state \(\j\) with \(a(\j)\ne 0\), has an infinite expectation value of this hamiltonian. If we would only be interested in all those states for which the hamiltonian is finite, these will automatically obey \(a(\j)=0\). So, for these ``physical" states, we can also ignore the non-canonical term in the \([q,p]\) commutator (the \(-1\) in Eq.~\eqn{pqcomm}).

We now see why also our harmonic oscillator has a divergent term in its hamiltonian, Eqs.~\eqn{qqelements}, \eqn{ppelements}, proportional to \(a(\j)\).
It is not difficult to generalize our result for hamiltonians that contain a magnetic term: \(H=\fract 1{2m}(p-{\cal A}(q))^2+V(q)\).

Clearly, all physical states (those states that have finite energy in the canonical \((q,p)\) system) have the property that, in \((\eta_Q,\eta_P)\) space, they vanish at the points in the corners: \(\eta_Q=\pm\half,\ \eta_P=\pm\half\). This is a non-trivial property; it is at these corners where the vector potential field \((a_Q,a_P)\) has a singularity.

Notice furthermore, that our argument would not be so straightforward if we had edge states for all values of either \(P\) or \(Q\) (that is, a singularity at the entire boundary of \((\eta_Q,\eta_P)\) space, instead of only at the corners), which is what our first attempts were giving us (see Eq.~\eqn{etaQcomm}). It is much better to have just one edge state.

\newsecl{Discussion}{disc}
	The study carried out in this paper could be viewed as a simple exercise in the application of matrices to transform from one basis of Hilbert space to another. However, it was done with a very special purpose, which is to demonstrate that quantum theories based on real numbers, such as the popsitions and momenta of elementary particles, or equivalently, the real values of bosonic field variables, can be mapped onto theories where the fundamental degrees of freedom are integers. Our aim is to apply this to cellular automaton theories of Nature, possibly of importance at or near the Planck scale. These theories 
would first have to be cast in the form of one of the more familiar quantum field theories by means of our transformation matrices, after which the renormalization group can be applied to determine their behavior at very large distance and low energy scales.	
	
	Of course, one would primarily think of \emph{quantum} theories acting on integer-valued oprators, but the author suspects that one may even go one (important) step further. The deterministic evolution of a classical cellular automaton may be cast in a mathematical framework where Hilbert space is introduced artificially\cite{GtHCA}, just to describe the evolution process as time proceeds. Only at integral time steps, this evolution would take the form of pure permutations in our discrete basis elements; due to the technique of introducing Hilbert space however, one can turn the time variable into a continuous variable of the kind that we think we experience in the physical world. Allowing ourselves the complete set of Hilbert space transformations to any basis we like, makes these theories mathematically hardly distinguishable from conventional quantum mechanics. Indeed, after applying these techniques, distinguishing deterministic theories from quantum mechanics may be so difficult at larger scales that this could be an explanation as to \emph{why} we think our world is quantum mechanical, while it may actually \emph{be} deterministic at the Planck scale. 
	
	We are aware of the fact that such arguments seem to be completely irreconcilable with the numerous examples of experimentally realizable situations\cite{epr} where \emph{Bell's inequalities}\cite{Bell}  are violated. To analyze exhaustively how this apparent disagreement can be resolved is beyond the aim of this paper (see for instance Ref.~\cite{GtHcollapse}). We would just emphasize the following facts:
	\bi {-} Our quantum states are real quantum states, including Born's identification of amplitudes squared with probabilities. There is nothing against the use of states of this kind that are as entangled as in real or imagined Aspect-like experiments, even if the underlying theory happens to be a deterministic one. 
	\itm{-} When considering Bell's inequalities in a deterministic system, in the real world, this system cannot be in just any quantum state; it is in exactly one of the fundamental cellular automaton states. In the real world no superpositions of such states can occur.	
	\itm{-} As stated however, a complete analysis is not possible without more detailed models. These we do not show here, but more will come in our next paper. \ei
\noindent In Section~\ref{harm}, a  brief illustration was given of a simple deterministic automaton. This particular example turns back to its original position after just four steps. The importance of this model is not the demonstration that it exists, since such models are quite trivial; the hamiltonian \(H\) merely needs to obey the condition that 
	\be \E^{-4iH}=\Bbb{I}\ ;\qquad H=\quart N +\ \hbox{integer}\ , \eel{harmham}
where the value of the integer is formally immaterial.

The importance of this model  is that we gave a very special representation of the hamiltonian, where the integer takes values such that we have a real harmonic oscillator. In particular, there is a natural ground state. This is important because it allows us to do thermodynamics with this oscillator. In our previous work, the choice of a good hamiltonian, with a non-trivial and physically meaningful ground state, has been problematic.

Now what we have is a \emph{mapping} of models defined in a discrete world, onto models that are quantum mechanical in the conventional sense, with \(p_i\) and \(q_i\) operators obeying 
\([q_i,p_j]=i\d_{ij}\). It sould be emphasized that there are different ways to perform this mapping, some of which may seem to be easier mathematically and conceptually, but we think that the procedure we ended up with is unique and superior. It is totally symmetric in \(p\) and \(q\), it has only one edge state, and, in spite of the rather awkward mathematical expression for the template states \(\bra 0,0|q\ket\) (Appendix~\ref{prop}), we ended up in remarkably simple expressions for the \(q\) and \(p\) matrix elements in \((Q,P)\) space (see Appendix~\ref{matrix}), so that the whole scheme ends up in being quite transparent.

Now our real aim is to apply this mapping in the case that the model in \((Q,P)\) space is a deterministic one. This means that, at integral time intervals \(t\) (if the fundamental time unit is normalized to 1), the evolution \(U(t)\) is a pure permutation (possibly a quite complicated one) on the \((Q,P)\) lattice. Although the observables are only well-defined and ``ontological" at integral time intervals, it is relatively easy to devise a hamiltonian \(H\) 
such that 
	\be U(t)=\E^{-iHt}\ . \eel{discrevolv}
This hamiltonian then, allows us to extrapolate the time variable to be continuous, but we think it is reasonable to postulate that, if the time quantum is sufficiently rapid, a ``physical observer" cannot distinguish between integer time states and non-integer time states.

It is an essential feature that, if we would add arbitrary integers to any of the energy eigenvalues, the original discrete model will not be affected at all, whereas the canonical \(q,p\) model may seem to become totally different. This means that, although our mapping \((Q,P)\leftrightarrow(q,p)\) is unique and inversible, there are many canonical hamiltonians that all correspond to the same discrete system. This freedom may actually be used to seek for a hamiltonian that is as close as possible to one of the standard quantum mechanical systems, such as a quantum field theory.

Work is under way on a sequel of this paper, where we construct such a physically much more interesting model: a quantum field theory in one space- and one time dimension, having a non-trivial hamiltonian with a ground state; we suspect that the significance of the present procedures will then become more evident than it may be now.

\appendix
 
\newsecl{Properties of the function \(\bra 0,0|q\ket\).}{prop}
 	To go from the real numbers \(q\) and \(p\) to the integers \(Q\) and \(P\), we use the real function defined in Eqs.~\eqn{pqphi} and \eqn{phiq},
		\be \j(q)=\j(-q)=\bra 0,0|q\ket=\bra q|0,0\ket\ , \eel{genfn}
which is equal to its own Fourier transform,
		\be \j(p)=\int\dd q\bra q|0,0\ket\E^{-i\,pq}\ . \eel{pqfourier}
The other matrix elements are simply given by
		\be \bra Q,P|q\ket=\j(q-Q)\,\E^{-iPq}\ . \eel{PQq}
 \(\j(q)\) is sketched in Fig.~\ref{states.fig}, and it plays a central role in our mappings. Because of the periodicity properties~\eqn{pseudoperiods}, the definition of the function \(\j\) can be written as
 		\be\j(x)=\int_{-\half}^\half\dd\eta\,\E^{i\f(\eta,x)}\ , \eel{psidef}
and the function \(\f(\eta,x)\) is given by Eq.~\eqn{phidef}, or
		\be r(\eta,x)\,\E^{i\f(\eta,x)}=\sum_{K=-\infty}^\infty\E^{-\half K^2+K(x+i\eta)}\ ;\qquad r,\ \f\ \hbox{ real.} \eel{phidef2}
This sum is a special case of the elliptic function  \(\vartheta_3\), and it can also be written as a product:
 		\be r(\eta,x)\,\E^{i\f(\eta,x)}=\prod_{K=1}^\infty(1-\E^{-K})\ \prod_{K=0}^\infty(1+\E^{x+i\eta-K-\half})(1+\E^{-x-i\eta-K-\half})\ , \eel{theta}
with \(r\) and \(\f\) real. Here, the first product term is of lesser importance since it only multiplies \(r(\eta,x)\) with a constant, while not contributing to \(\f(\eta,x)\). Note, that  \(\f(\eta,x)\) has a vortex singularity when \(\E^{i\f(\eta,x)}\) has a zero, and these zeros can easily be read off from Eq.~\eqn{theta}; they are located at \((\eta,\,x)=(K_1+\half,\,K_2+\half)\).
We see that in Eq.~\eqn{psidef}, the absolute value \(r(\eta,x)\) of the sum in \eqn{phidef2}, or the product in \eqn{theta}, has been divided out, and this makes the evaluation of the integral over \(\eta\) hard, although it is well bounded. 
 
In Eq.~\eqn{phidef2}, the sum is dominated by the \(K\) value closest to \(x\). In Fig.~\ref{states.fig}, the small peaks at large \(x\) (Fig.~\ref{states.fig}$e$) arise when the dominant \(K\) value in the sum switches from one integer to the next.
 
\emph{Unitarity property}\,:\\
From the fact that \(\E^{i\f(\eta,x)}\) in Eq.~\eqn{psidef} is the Fourier transform of \(\j(x)\) for integral \(p\), and that it has absolute value one, we derive that
\be\sum_{K=-\infty}^\infty \j(x+K)\,\j(x+K+M)=\d_{M\,0}\ ,\qquad K,L\in\Bbb Z\ \eel{unitar} 
 (use was made of Eq.~\eqn{phiq}).
 
\newsecl{Matrix elements of \(q\) and \(p\) operators}{matrix}

The  matrix elements \(\bra Q_1,P_1|\,q\,|Q_2,P_2\ket\) can be calculated explicitly. Let us first compute the operator \(q\) in \(\eta_Q,\eta_P\)-space. We write the expression \eqn{qopetaspace} as follows:
	\be q={-i\over 2\pi}{\pa\over\pa \eta_Q}+a_Q(\eta_Q,\eta_P)\ , \qquad a_Q(\eta_Q,\eta_P)={\pa\f(\eta_P,\eta_Q)\over\pa\eta_Q}\ , \eel{qopcov}
where the function \(a_Q\) is regarded as the \(Q\)-component of a vector potential field \(a\). 
For the phase \(\f(\eta_P,\eta_Q)\), we can now best use the product formula \eqn{theta}, which gives:
	\be a_Q(\eta_Q,\eta_P)&=&\sum_{K=0}^\infty a_Q^K(\eta_Q,\eta_P)\ ,\crl{aQK}  
	a_Q^K(\eta_Q,\eta_P)&=&{\pa\over 2\pi\,\pa  \eta_Q}\bigg(\arg(1+\E^{\eta_Q+i\eta_P-K-\half})+\arg(1+\E^{-\eta_Q-i\eta_P-K-\half})\bigg)\ . \eel{aQKarg}
Evaluation gives:
	\be &&a_Q^K(\eta_Q,\eta_P)\ = \cr &&
	{\half\sin(2\pi\eta_P)\over\cos(2\pi\eta_P)+\cosh(2\pi(\eta_Q-K-\half))}
	 +{\half\sin(2\pi\eta_P)\over\cos(2\pi\eta_P)+\cosh(2\pi(\eta_Q+K+\half))}\ ,\qquad{\ }	\eel{aQexpr}
which can now be rewritten in a more compact way by rewriting Eq.~\eqn{aQK}  as a sum for \(K\) values running from \(-\infty\) to \(\infty\) instead of \(0\) to \(\infty\).

By writing 
	\be\bra P_1|\eta_P\ket\bra\eta_P|P_2\ket=\E^{iP}\ ,\qquad P\equiv P_2-P_1\ , \eel{Petabasis}
we now proceed to write the matrix elements of the operator \(a_Q\) in the \((\eta_Q,\,P)\) frame: 
	\be \bra P_1|a_Q(\eta_Q)|P_2\ket=\sum_{K=-\infty}^\infty a_Q(\eta_Q,\,P,\,K)\ ,\qquad P\equiv P_2-P_1\ , \ee
finding
	\be a_Q(\eta_Q,\,P,\,K)=\half\sgn(P)(-1)^{P-1}i\E^{-\big|\,P(\eta_Q+K+\half)\,\big|}\ ,\eel{aQPK}
where \ sgn\(\!(P)\) is defined to be \(\pm 1\) if \(P\gl 0\) and \,0\, if \(P=0\). The absolute value taken in the exponent indeed means that we always have a negative exponent there; it originated when the contour integral forced us to choose a pole inside the unit circle.

Next, we find the \((Q,P)\) matrix elements by integrating this with a factor \(\E^{iQ}\), with \(Q=Q_2-Q_1\), to obtain the remarkably simple expression
	\be\bra Q_1,P_1|a_Q|Q_2,P_2\ket\iss{(-1)^{P+Q+1}\,iP\over P^2+Q^2}\  .\eel{aQmatrix}
In Eq.~\eqn{qopcov} this gives for the \(q\) operator:
	\be q=Q+a_Q\ ;\qquad \bra Q_1,P_1|q|Q_2,P_2\ket=Q_1\d_{Q_1\,Q_2}\,\d_{P_1\,P_2}+\bra Q_1,P_1|a_Q|Q_2,P_2\ket\ . \eel{qmatrix}
	
For the \(p\) operator, one obtains analogously, writing \(P\equiv P_2-P_1\),
	\be  &p=P+a_P\ ,& \crl{pmatrix}
	&\bra Q_1,P_1|a_P|Q_2,P_2\ket &=\ {(-1)^{P+Q}\,iQ\over P^2+Q^2}\  . \eel{aPmatrix}
	
It is important to check the commutation rule for \(q\) and \(p\). Doing the matrix multiplications for the matrices \eqn{qmatrix} and \eqn{pmatrix}, one finds that
	\be &[Q,P]=0\ ,\qquad[a_Q,\,a_P]=0\ ,& \crl{PQaacomm}
	&\bra Q_1,P_1|[q,p]|Q_2,P_2\ket\iss\bra Q_1,P_1|\,[Q,\,a_P]+[a_Q,\,P]\,|Q_2,P_2\ket\ =&\nm \\[5pt] 
	&i\,(-1)^{Q_1-Q_2+P_1-P_2}(\d_{Q_1\,Q_2}\d_{P_1\,P_2}\ -\,1)\ .& \eel{pqcomm}
Again, we see that the desired commutation rule, \([q,p]=i\), is obeyed only after we project out the edge state \(\j_\edge\) by demanding that all our states must obey
	\(\bra\j_\edge|\,\j\,\ket=0\), see Eq.~\eqn{singleedge}, see Section~\ref{edge}.

\end{document}

If truly entangled quantum states emerge in a model that is deterministic at some tiny scale such as the Planck scale, that does not contradict anything. We do have to remember that if a quantum state is entangled now, it must have been entangled in the far past as well, so this observation seems to suggest that, \emph{in terms of standard model variables}, the world has always been in an entangled state. In terms of the original, ``ontological" basis of the deterministic variables, one could suspect that we have an entangled state as well, but this entanglement would most likely have to be attributed to our lack of exact knowledge of the initial state, so that, \emph{in our perception}, it appears to be in a mixed state.

 and this is why, in an Aspect-like, or in an EPR experiment, it is not allowed to consider rotations of a measurement device in such a way that quantum superpositions of operators are measured; if one attempts to do this anyway, for instance by rotating Bob's measurement device, it might affect the (not directly observable) quantum phases of the states Alice has observed, otherwise one cannot have an ontological state in the automaton.